# The Influence of Culture on Migration Patterns

Tomáš Evan[1]

Czech Technical University in Prague

Thákurova 2077/7, 160 00 Prague 6, Czech Republic

University of New York in Prague

Londýnská 41, 120 00 Prague 2, Czech Republic

tomas.evan@fit.cvut.cz

Eva Fišerová

Palacký University, Olomouc, Faculty of Science

17. listopadu 12, 771 46 Olomouc, Czech Republic

eva.fiserova@upol.cz

Aneta Elgnerová

Palacký University, Olomouc, Faculty of Science

17. listopadu 12, 771 46 Olomouc, Czech Republic

aneta.elgnerova@upol.cz

---

[1] Corresponding author

# Abstract

UN migration data and Hofstede's six cultural dimensions make it possible to find a connection between migration patterns and culture from a long-term perspective. Migrant patterns have been studied from the perspective of both immigrants and OECD host countries. This study tests two hypotheses: first, that the number of migrants leaving for OECD countries is influenced by cultural similarities to the host country; and second, that OECD host countries are more likely to accept culturally close migrants. Both hypotheses were tested using the Mann–Whitney U test for 93 countries between 1995 and 2015. The relationship between cultural and geodesic distance also analysed. The results indicate that cultural proximity significantly influences migration patterns, although the impact varies across countries. About two-thirds of OECD countries show a positive correlation between cultural similarity and geographic proximity, with notable exceptions, such as New Zealand and Australia, which exhibit a negative correlation. Countries such as Colombia, Denmark, and Japan maintain cultural distance, even from their neighboring countries. Migrants from wealthier countries tend to select culturally similar destinations, whereas those from poorer countries often migrate to culturally distant destinations. Approximately half of OECD countries demonstrate a statistically significant bias towards accepting culturally close migrants. The results of this study highlight the importance of a critical debate that recognises and accepts the influence of culture on migration patterns.

# 1 Introduction

International migration, along with the connected topics of aging and depopulation in developed countries, has become one of the most pressing topics today. Half of the countries in the world, and almost all developed nations, have a total fertility rate below the threshold of 2.1 children per woman needed for a sustainable population according to the United Nations [UN] 2019 Revision of World Population Prospects. The decline in mortality in the last two centuries has mostly levelled out. Government policies aimed at facilitating women's reconciliation of work and family life appear unable to mitigate the negative impact on the natality of a large share of women in the workplace. The opportunity costs of the decreased wage gap between males and females also decrease natality, as do the many social benefits of the modern welfare state (Evan and Vozárová, 2018). This lack of people and youth, in particular, has a profound impact on virtually every segment of modern society, from military capabilities and research expertise to real estate, retail markets, and household indebtedness. As natural reproduction levels are out of reach, large, developed countries need hundreds of thousands of additional people per annum to maintain their workforce. In turn, this places importance on the migration policies of countries to find the most suitable prospective citizens (Bauböck et al., 2022; Dumitru, 2023).

It is accepted that economic drivers of migration are the most potent. However, since economic conditions are constantly changing, we would like to focus on culture's influence on migration, as its relative immutability makes it a significant long-term factor. Recently, a significant body of literature has focused on cultural diversity. It has become clear that there are some significant disadvantages connected with it, *inter alia* political instability, decreased economic productivity, and less efficient governance (for details, see 2 Theoretical Background).

Despite the low or even possibly non-existent assimilation in some countries, the prime source of increased cultural diversity is ongoing immigration, as there is likely no other way to sustain or increase cultural diversity. In our study, we want to determine whether these facts influence the current migration policies of the majority of developed countries (i.e., Organisation for Economic Co-operation and Development [OECD] countries), which serve as host countries for much of the existing world migration.

We propose a model in which we study the long-term dependence of average migration flows taken from the UN database (UN, 2015) on the cultural similarities between countries. We used Hofstede's six cultural dimensions to define cultural distances among countries (Hofstede, 2010). This allows us to assess how many OECD countries with different approaches to immigration are inclined to accept migrants based on cultural similarity and/or how many migrants take culture into consideration when choosing a destination (1). We also aimed to disentangle the relationship between cultural and geodesic distance to ensure that cultural distance is not simply a direct consequence of geodesic distance (2). Finally, we attempt to quantify how similarities in national cultures influence migration overall and how strong and significant the relationship is (3).

The remainder of this paper is organised as follows. In Section 2, we summarise Hofstede's six-dimensional cultural model and discuss cultural stability, immigration patterns, and cultural perseverance. Section 3 presents the model and data sample. Section 4 describes and interprets the results. We conclude the paper in Section 5.

# 2 Theoretical Background

2.1 Hofstede's Cultural Dimensions

Hofstede's six cultural dimensions make it possible to measure differences in national culture. Hofstede et al. (2010) defined the six cultural dimensions as follows: Power Distance (i) is the extent to which less powerful members of institutions and organizations within a country expect and accept that power is distributed unequally. Individualism (ii) pertains to societies in which the ties between individuals are loose, as everyone is expected to look after themselves and their immediate family. Masculinity (iii) refers to a society in which emotional gender roles are clearly distinct: men are supposed to be assertive, tough, and focused on material success, whereas women are supposed to be more modest, tender, and concerned with quality of life. Uncertainty Avoidance (iv) is the extent to which members of a culture feel threatened by ambiguous or unknown situations. Long-Term Orientation (v) refers to the fostering of virtues oriented towards future rewards, in particular, perseverance and thrift. Finally, Indulgence (vi) indicates a tendency to allow relatively free gratification of basic and natural human desires related to enjoying life and having fun. (See the website www.hofstede-insights.com/country-comparison for a detailed description of the six cultural dimensions.)

2.2 Cultural Stability

The key part of our analysis depends on the virtual immutability of culture, as described by Hofstede (Hofstede, 1980, 1994, 2001). If there are other means of cultural change beyond the immigration of individuals from different cultural backgrounds, the importance of immigration policies and the decisions of prospective migrants about their destinations would be greatly reduced.

Theoretical predictions have been made not only for stability but also for the convergence of cultures (Webber, 1969; Ralston et al., 1997; Inglehart and Baker, 2000; Leung et al., 2005; McSweeney, 2009; Tung and Verbeke, 2010, i.a.). The continuous use of Hofstede's four, five, and ultimately six cultural dimensions allowed the authors to test the hypothesis using data over several decades. Beugelsdijk et al. (2013), in a paper called "Are Hofstede's Culture Dimensions Stable Over Time? A Generational Cohort Analysis", are among those suggesting that cultures generally are stable. The overall conclusion includes claims about the replicability of national culture for different cohorts and, equally important, stable average cultural distance (with the exception of the USA) among countries (Beugelsdijk, 2013). However, they also reported small changes in three of the six dimensions across the board, namely increased levels of Individualism and Indulgence, while they observed lower scores for the Power Distance dimension. This is in line with Tang and Koveos (2008), who claim that dramatic changes in economic conditions can be the source of minor cultural change. They found that Individualism, Long-Term Orientation, and Power Distance are less permanent than the Uncertainty Avoidance and Masculinity dimensions. These results were confirmed over a longer timeframe by Zhao et al. (2016) and Matei and Abrudan (2018). The stability and relevance of this 50-year model can be considered proven by the caveats of small changes in several cultural dimensions under specific conditions. These findings allow us to use cultural distance in our analysis, as rapid economic development spread over an extended period does not alter the pertinence of cultural distance in our study. The differences between dimensions with and without these minor changes over time should be examined in future research.

The literature has already established that culture is very stable over time, but for the overall influence of culture on migration patterns, it is equally important whether it is geographically stable – that is, whether holders of one culture keep it regardless of where they immigrate. This is also a reasonable assumption. To provide just one example, Alnunu et al. (2021) tested this by examining Arab Muslims in the Arab region, those who underwent Western cultural socialization, and those who migrated to East Asian cultures. They found that the cultural patterns of susceptibility to persuasion were the same regardless of the host country and the acculturation model (assimilation, integration, separation, and marginalization).

2.3 Cultural patterns in immigration

Large immigration can drastically change the cultural, institutional, political, and socioeconomic patterns of the host country. Equally significant is the change in the source regions of migration flows to the host countries. If we take the USA as an example, 70.4% of the immigrants into the country in 1970 were from North America and Europe, while in 2019, 50.3% were from Latin America and 31.4% from Asia, and only 12.2% (almost six times less) were from North America and Europe (US Census Bureau 2019: ACS 1-Year Estimates Data Profiles). Such a rapid shift in the source of migrating cultural groups is likely to profoundly impact social and institutional backgrounds such as social stratification and mobility (Massey, 1995). As others have shown (Beugelsdijk, 2013), the U.S. immigration policy over the last 50 years is unique. It shows a certain arbitrariness, regardless of economic development as well as the culture of immigrants. As stated in the introduction, our task in this text is to determine whether the cultural distance of the potential host country and the reaction to the fact by both policymakers and potential migrants can indeed influence the number of immigrants.

We suggest that immigrants generally choose a culturally close country to avoid adjusting to a different socioeconomic environment and to increase their chances of economic success in host societies. In further research, it might be interesting to prove with any statistical relevance whether pre-existing cultural communities influence international migration flows as well as the areas within the host countries where migrants choose to reside. As such, the cultural differences we examined may effectively decrease the cost of immigration through existing social networks (Manchin and Orazbayev, 2018). Regardless, as culture is slow to change and even migrants tend to retain their culture, cultural diversity increases, irrespective of where they decide to settle. This is claimed to cause instability and internal conflict (Easterly, 2001; Nettle et al., 2007, Kónya, 2007, i.a.), diminishing trust among different groups and decreasing the formation of social capital (Taylor, 2000; Zak and Knack, 2001; Greve and Salaff, 2003; Putnam, 2007, i.a.). Additionally, it is argued that it causes weak economic performance in the host country, as measured by either growth or productivity (Nettle, 2000; Grafton et al., 2002; Alesina et al., 2003, i.a.). A summary of the literature on the impact of high cultural diversity on socioeconomic variables and general well-being can be found in Evan and Holy (2023).

Moreover, the immutability of culture means that migrants from culturally closer countries would cause less friction in the host society, as cultural differences are significant and influential. Culture is involved in everyday decisions either directly or through institutions; a "frozen" or "formalized" culture (Olson, 1996). Relevant literature attests to the difficulties associated with the coexistence and cooperation of people with different cultures (Williams and O'Reilly III, 1998; Earley and Mosakowski, 2000; Richard et al., 2004, i.a.). Even the most open-minded people must reflect the different patterns of thinking shaped by language and religious axioms, both of which underlie the values of the six cultural dimensions. This leads to a cautious approach and less communication, which hinders social interaction, particularly if there is an experience of previous misunderstandings (Østergaard et al., 2011; Harvey and Kou, 2013). Ignoring these well-established facts when setting up the parameters of any migration policy can be considered irresponsible.

Cultural assimilation may be required from immigration policy, depending on the willingness and ability of the majority culture to adapt (Leong and Ward, 2006). This is certainly a way to reduce cultural diversity, as is often preferred by the majority culture. In this case, the task of the immigration policy is to admit only a small number of culturally similar migrants, thus creating a society in which all important values and institutions are unified. As a result, people of different races, ethnicities, and even religions, which are initially culturally distinct, are ultimately holders of the same culture. The other option is cultural divergence where, to the extent of their numbers and abilities in particular social areas, immigrant cultures contribute some of their own values to the constantly changing or blended majority culture. Even more adaptation by all members of society is needed for hypothetical cultural pluralism (multiculturalism), where cultures live in their own space, each remaining clearly identifiable as only economic integration exists. When large numbers of immigrating people from different cultures are accepted, this last possibility, sometimes called the Balkanization of society (Brooks, 2012), is an outcome that the host society should expect.

2.4 Perseverance of culture

Economic reasons for migration have long been considered the most significant. Standard economic theory traditionally has the power to explain migration levels and can even measure the benefits of migration to particular groups within society. According to any regular textbook on international economics, when economic migrants move from a country with abundant labour to a country with abundant capital, the economic outcome is positive. Migrants benefit, as do the capitalists in the host country and labour in the home country, while higher world production is achieved. However, these models fail to account for differences between people other than the standard ones, such as education level and age. This is highly inadequate, as direct costs associated with migration, intergroup contact, exchange costs, or cultural transaction costs in the host country (Newman et al., 2014) may be substantial enough to limit or even negate the benefits described by the economic model. While it might be possible to expand the economic analysis to include the direct costs of migration, discerning the psychological damage would be far more difficult. Most importantly, economic concepts cannot account for the loss of social capital incurred by anyone migrating to a culturally distinct country.

Moreover, given economic realities such as wage or unemployment differentials, the question is: why are there not significantly more migrating people in the world? In Europe, a migrating European has been called "the endangered bird" (Braunerhjelm et al. 2000), as migration within Europe declined at the end of the last millennium, despite the persistent economic disparities. Although we do not fully understand the relative influence of economic factors on migration, which sometimes seems quite paradoxical (Czaika and de Haas, 2012), the main culprit appears to be the cultural differences that significantly and perhaps prohibitively increase the costs of such migration. Belot and Ederveen (2012) tested the effect of cultural barriers on migration flows among 22 OECD countries from 1990 to 2003, and found that culture explains migration flows better than income and unemployment differentials.

In the case of one country in particular, Germany, Bauernschuster et al. (2013) also found that cultural patterns (risk adversity) as well as the level of education affect people's long-distance mobility. Using data from the 1879 and 1888 language surveys showing local German dialect patterns, they found that more skilled and less risk-averse people are willing to cross regional cultural boundaries and move to destinations that are more culturally different than those lacking those qualities. Falck et al. (2012) used the same data to find that these 140-year-old cultural barriers were still visible in German cross-regional migration flows between 2000 and 2006. Thus, they demonstrated that cultural ties are highly persistent over time, even in inland migration. Furthermore, using the gravity model, Lanati and Venturini (2021) establish an international migration model in which cultural distance is not symmetric. This model allows for positive changes in the cultural relations between countries, such as trade in cultural goods, to foster bilateral migration.

# 3 Data and methods

Hofstede's data for six dimensions of culture (Hofstede, 2010) were used in the analysis. Dimension scores were scaled from 0 to 100. The estimates for missing values and data for several other countries were obtained from Hofstede Insight's Cultural Executive Ownership Program website. (See the website www.hofstede-insights.com/country-comparison for details.) Data on all six cultural dimensions were obtained for 93 countries, including 36 OECD countries. Israel was the only OECD country excluded from the study due to incomplete cultural data.[2]

Countries can be directly compared based on differences in any two cultural dimensions. A country comparison graph is a useful tool that enables users to visually analyse differences in up to three dimensions among any combination of countries. It presents the data as a two-dimensional scatter plot, with colour shading representing an additional third dimension (for details, see https://geerthofstede.com/country-comparison-graphs). To quantitatively analyse the overall cultural similarity of countries, we define the *cultural distance*, *CD*, between two countries $C_1$ and $C_2$ with respective cultural dimensions $pdi_1$ (power distance index), $idv_1$ (individualism versus collectivism), $mas_1$ (masculinity versus femininity), $uai_1$ (uncertainty avoidance index), $ltowvs_1$ (long-term orientation versus short-term normative orientation), $ivr_1$ (indulgence versus restraint), and $pdi_2, idv_2, mas_2, uai_2, ltowvs_2, ivr_2,$ as the Euclidean distance with the component differences.

---

[2] A certain imprecision in power distance (pdi) and masculinity (mas) equal to 100 for Slovakia was found in the dataset. Based on a brief communication with Gert Jan Hofstede, *pdi* was set at 80 and *mas* at 60 for this country.

$$\Delta pdi_{C_1,C_2} = pdi_1 - pdi_2, \qquad \Delta idv_{C_1,C_2} = idv_1 - idv_2, \qquad \Delta mas_{C_1,C_2} = mas_1 - mas_2,$$

$$\Delta uai_{C_1,C_2} = uai_1 - uai_2, \qquad \Delta ltows_{C_1,C_2} = ltows_1 - ltows_2, \qquad \Delta ivr_{C_1,C_2} = ivr_1 - ivr_2$$

$$CD(C_1, C_2) = \sqrt{\Delta pdi_{C_1,C_2}^2 + \Delta idv_{C_1,C_2}^2 + \Delta mas_{C_1,C_2}^2 + \Delta uai_{C_1,C_2}^2 + \Delta ltows_{C_1,C_2}^2 + \Delta ivr_{C_1,C_2}^2}$$

Cultural distance between the two countries ranged from 0 to 245. A lower cultural distance indicates a higher overall cultural similarity between these countries and vice versa. All pairs of countries were further stratified into one of three categories according to cultural distance: culturally close, culturally mid-distant, or culturally distant. The thresholds for these categories are chosen as the lower (*CD* upper limit for culturally close countries) and upper quartile (*CD* lower limit for culturally distant countries) of pairwise cultural distances among all analysed countries.

Data on international migration flows among all 93 countries were obtained from the UN database (UN, 2015). The UN International Migration Flows database contains only aggregated annual data on the flows of international migrants to and from particular countries, without specifying their country of origin or destination. Therefore, the International Migrant Stock database was used to estimate the average annual migration flows between country pairs from 1995 to 2015. For comparability, average annual migration flows were normalised per 1 million inhabitants of the country of origin or the host country. The average annual population of all 93 countries between 1995 and 2015 was estimated using population data from the World Bank database (for details, see https://data.worldbank.org/indicator/SP.POP.TOTL).

We define the geodesic distance between two countries as the geodesic distance (Karney, 2013) between their capital. The distance is calculated as the shortest path between the two capitals over Earth's surface. To obtain the respective positions on the Earth's surface, we used the World Geodetic System 1984 (WGS84), which consists of a standard coordinate system and the best available ellipsoid model for Earth. The calculations were performed using the distGeo function in the R package geosphere (Brown et al., 2021).

To describe the location, spread, and scale of the data, we used robust measures: the median, lower/upper quartile, and median absolute deviation (MAD), respectively. Considering the nature of the data, a distribution-free methodology was applied for statistical analysis. The one-sided two-sample Mann–Whitney U test was used to test how: (i) the number of migrants coming to OECD countries between 1995 and 2015 from culturally close and culturally distant countries differ; (ii) the number of migrants leaving for OECD countries between 1995 and 2015 varies based on whether they are culturally close or culturally distant from their home country; and (iii) the geodesic distances of culturally close and distant countries to OECD countries differ. If there were less than or equal to three observations in the category of culturally distant/close countries, the category of culturally distant countries was used in the test procedure. The one-sided Wilcoxon signed-rank test was used to verify the significance of (i) migration to OECD countries; and (ii) leaving OECD countries of origin for OECD countries. For significant results visualised in figures and tables, the following symbols are used: *** denotes p-value < 0.001, ** p-value <0.01, * p-value < 0.05 and "black dot" p-value < 0.1. The open-source software environment R was used for computation (R Core Team, 2020).

# 4 Results

The dataset consists of 36 OECD countries (Israel is excluded for the incompleteness of cultural dimensions) and 57 non-OECD countries. Only OECD countries are considered as host countries. We have 2,052 observations for migration from non-OECD countries to OECD countries, and 1,260 observations for migration from OECD countries to other OECD countries. In total, we have 3,312 observations of average migration per year during 1995–2015.

The cultural and geodesic distances between the analysed countries are summarised in Table 1. The median cultural distance was $CD$=78.0. The culturally closest countries (see Table 2) are Australia and the USA, with $CD$=8.1, and Estonia and Lithuania, with $CD$=12.2. Note that while Estonia and Lithuania have very short geodesic distances (GD=530 km), Australia and the USA have very large geodesic distances (GD=15945 km). The most culturally distant countries (see Table 3) are Venezuela and Latvia ($CD$=139.2), and Ukraine and Denmark ($CD$=137.5). Again, Ukraine and Denmark are geographically close, and Venezuela and Latvia are geographically mid-distant. The lower quartile of cultural distance was 61.6, and the upper quartile was 93.4. If the cultural distance between the two countries was less than or equal to 61.6, the countries were classified as culturally close. If the cultural distance was greater than 93.4, they were classified as culturally distant. In other cases, they are classified as culturally mid-distant. For each OECD country, the remaining 92 countries (35 OECD countries and 57 non-OECD countries) were classified according to their cultural distance from the given country. There is clear inhomogeneity in the structure of cultural distances between pairs of OECD and non-OECD countries (see Figure 1). Australia, Austria, Denmark, the Netherlands, New Zealand, Sweden, the United Kingdom, and the USA are culturally distant from more than half of the non-OECD countries, and rarely are they culturally close to some non-OECD countries. In contrast, Greece, Chile, Poland, Portugal, Slovenia, Spain, and Turkey are culturally close to more than half of the non-OECD countries, and only rarely are they culturally distant from some non-OECD countries. Inhomogeneity in cultural distance is also visible between OECD countries (see Figure 2). In general, OECD countries are either culturally close or mid-distant. The Czech Republic, France, Luxembourg, and Switzerland are culturally close to more than half of OECD countries, and only rarely are they culturally distant from some OECD countries. The median geodesic distance between the considered pairs of countries is 6,347 km. The shortest geodesic distance ($GD$) is between Austria and Slovakia ($GD$=56 km) and Estonia and Finland ($GD$=83 km). In contrast, the largest geodesic distance is between New Zealand and Spain ($GD$=19851 km) and between Indonesia and Colombia ($GD$=19811 km). While Estonia and Finland are culturally close countries, the remaining three pairs of countries can be classified as culturally mid-distant.

|  | Min. | Lower quartile | Median | Mean | Upper quartile | Max. | MAD |
|---|---|---|---|---|---|---|---|
| Cultural distance | 8.06 | 61.58 | 78.02 | 77.23 | 93.38 | 139.20 | 23.32 |
| Geodesic distance [km] | 56 | 2,294 | 6,347 | 6,569 | 9,664 | 19,851 | 5,743 |

*Table 1: Summary of cultural and geodesic distances between analysed countries.*

| Country | Country | Geodesic distance [km] | Cultural distance |
|---|---|---|---|
| Australia | USA | 15,945 | 8.06 |
| Estonia | Lithuania | 530 | 12.24 |
| Brazil | Turkey | 10,360 | 14.53 |
| Iceland | Norway | 1,753 | 16.70 |
| Spain | Turkey | 3,092 | 17.78 |
| Canada | USA | 734 | 18.06 |
| Canada | New Zealand | 14,473 | 19.62 |

| Latvia | Lithuania | 264 | 19.65 |
| Greece | Turkey | 820 | 20.27 |
| Hungary | Italy | 812 | 20.37 |
| Australia | Canada | 16,104 | 20.62 |
| Venezuela | Mexico | 3,602 | 21.19 |

*Table 2: Pairs of culturally closest countries arranged in ascending order, and their geodesic distance.*

| Country | Country | Geodesic distance [km] | Cultural distance |
|---|---|---|---|
| Venezuela | Latvia | 9,096 | 139.20 |
| Ukraine | Denmark | 1,331 | 137.55 |
| Belarus | Denmark | 982 | 136.64 |
| Venezuela | Lithuania | 9,204 | 135.18 |
| Belarus | Ireland | 2,217 | 134.79 |
| Albania | Denmark | 1,681 | 134.56 |
| Kazakhstan | Denmark | 3,824 | 134.17 |
| Ukraine | Ireland | 2,523 | 133.59 |
| Belarus | Australia | 15,104 | 133.52 |
| Belarus | New Zealand | 17,215 | 133.08 |
| Ukraine | Australia | 14,887 | 133.07 |
| Cabo Verde | Japan | 14,135 | 132.96 |

*Table 3: Pairs of the most culturally distant countries arranged in descending order, and their geodesic distance.*

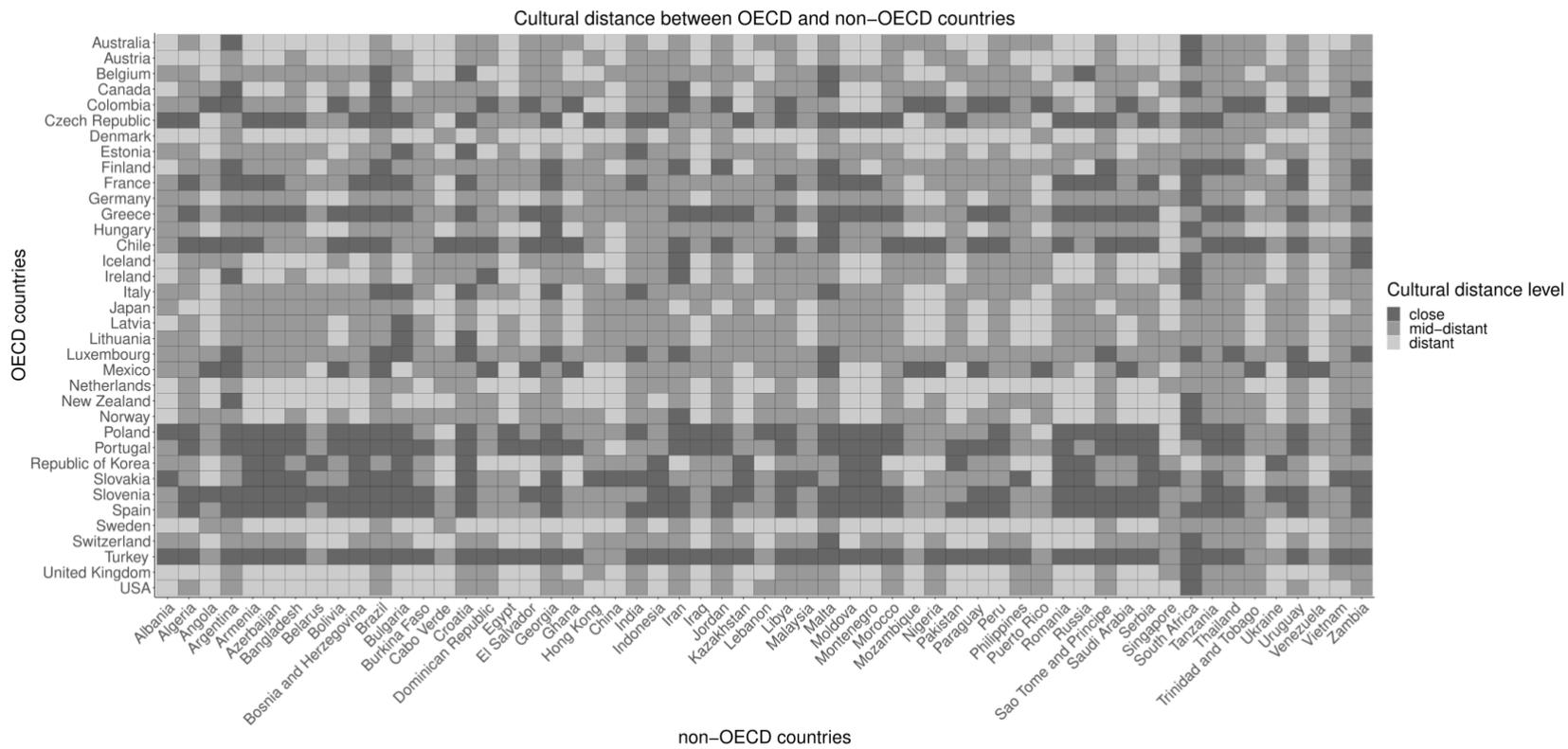

*Figure 1: Cultural distances between pairs of OECD and non-OECD countries. The pair of countries is classified as culturally close (distant) if the cultural distance is less than 61.6 (greater than 93.4); otherwise, the pair is considered culturally mid-distant.*

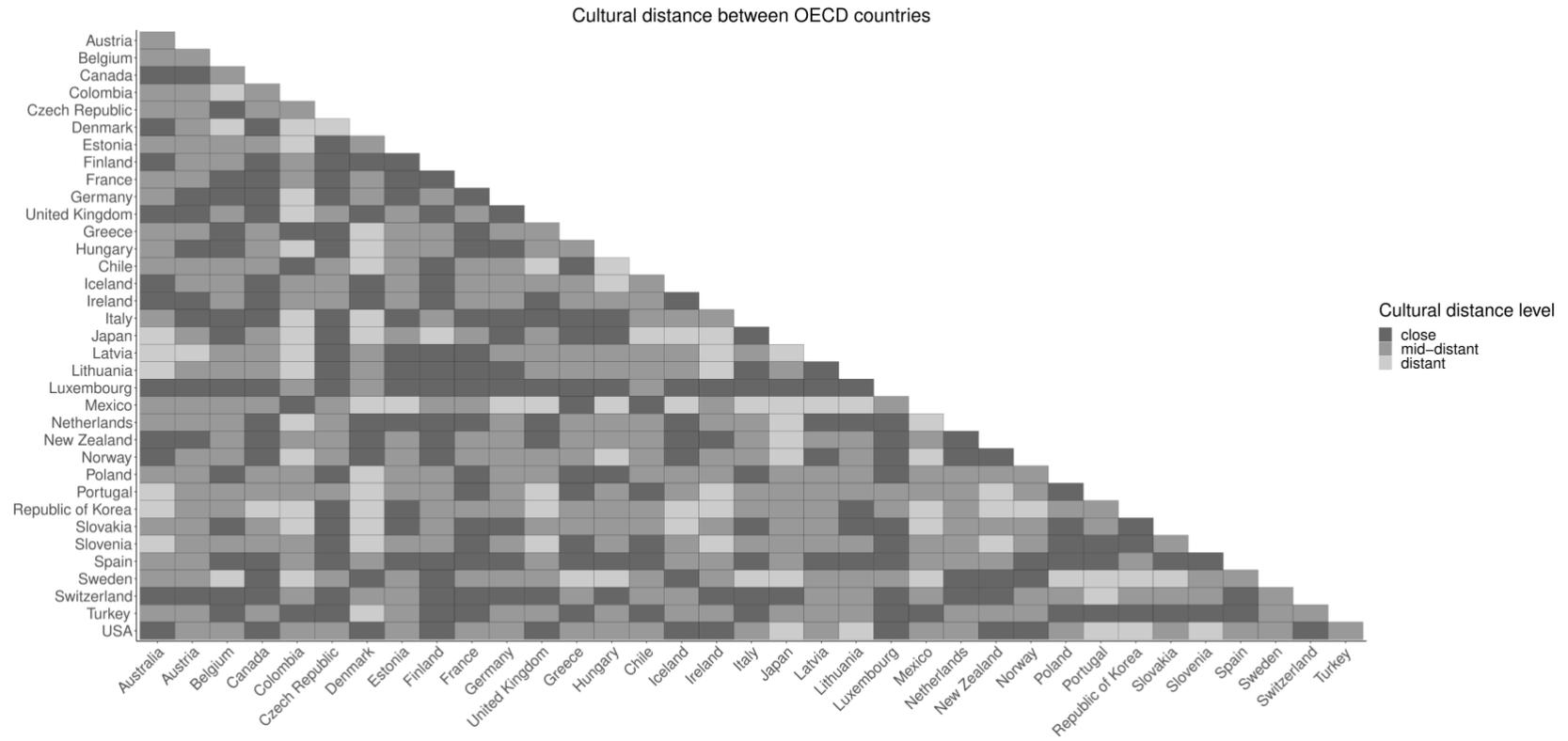

*Figure 2: Cultural distances between pairs of OECD countries. The pair of countries is classified as culturally close (distant), if the cultural distance is less than 61.6 (greater than 93.4); otherwise, the pair is considered culturally mid-distant.*

First, let us focus on the relationship between cultural and geodesic distance. Overall, the grand median of the geodesic distance from the considered countries to OECD countries was 6,340 km. In other words, the median of the shortest distances between pairs of capital with at least one OECD member country is 6,340 km. The medians of geodesic distances to OECD countries from culturally close, mid-distant, and distant countries were 3,227 km, 6,710 km, and 7,733 km, respectively. As expected, the median geodesic distance increased with increasing cultural distance, and vice versa. However, if we look at individual OECD countries in detail and compare their geodesic distance to culturally close and culturally distant countries, the results are no longer clear (see Figure 3). The trend of increasing geodesic distance with increasing cultural dissimilarity is significant for more than two-thirds of OECD countries (25 of 36 countries). The decreasing trend is significant only for New Zealand and Australia; that is, cultural differences decrease with increasing geodesic distance and vice versa. This phenomenon may be explained by the fact that both are former colonies of the United Kingdom, making them culturally close to it and its other major former colonies, such as the USA or Canada, despite their geodesic distance. For Portugal, Slovenia, Poland, Iceland, Ireland, Finland, Norway, Japan, and Turkey, there was no significant trend in geodesic distance with respect to cultural distance.

Now, let us examine the hypothesis that the average number of migrants coming to OECD countries is related to cultural similarities to the host country. In general, the grand median of average migration to OECD countries between 1995 and 2015 is 1.51 per year per 1 million inhabitants of the host country. The median average migration from culturally close countries is 1.54 per year per 1 million inhabitants of the host country. For culturally mid-distant countries, the median was 1.30, and for culturally distant countries it was 2.36. Half of the OECD countries (18 out of 36) show significant bias, whether positive or negative, towards cultural similarity between the migrant and the potential host country (see Figure 4).

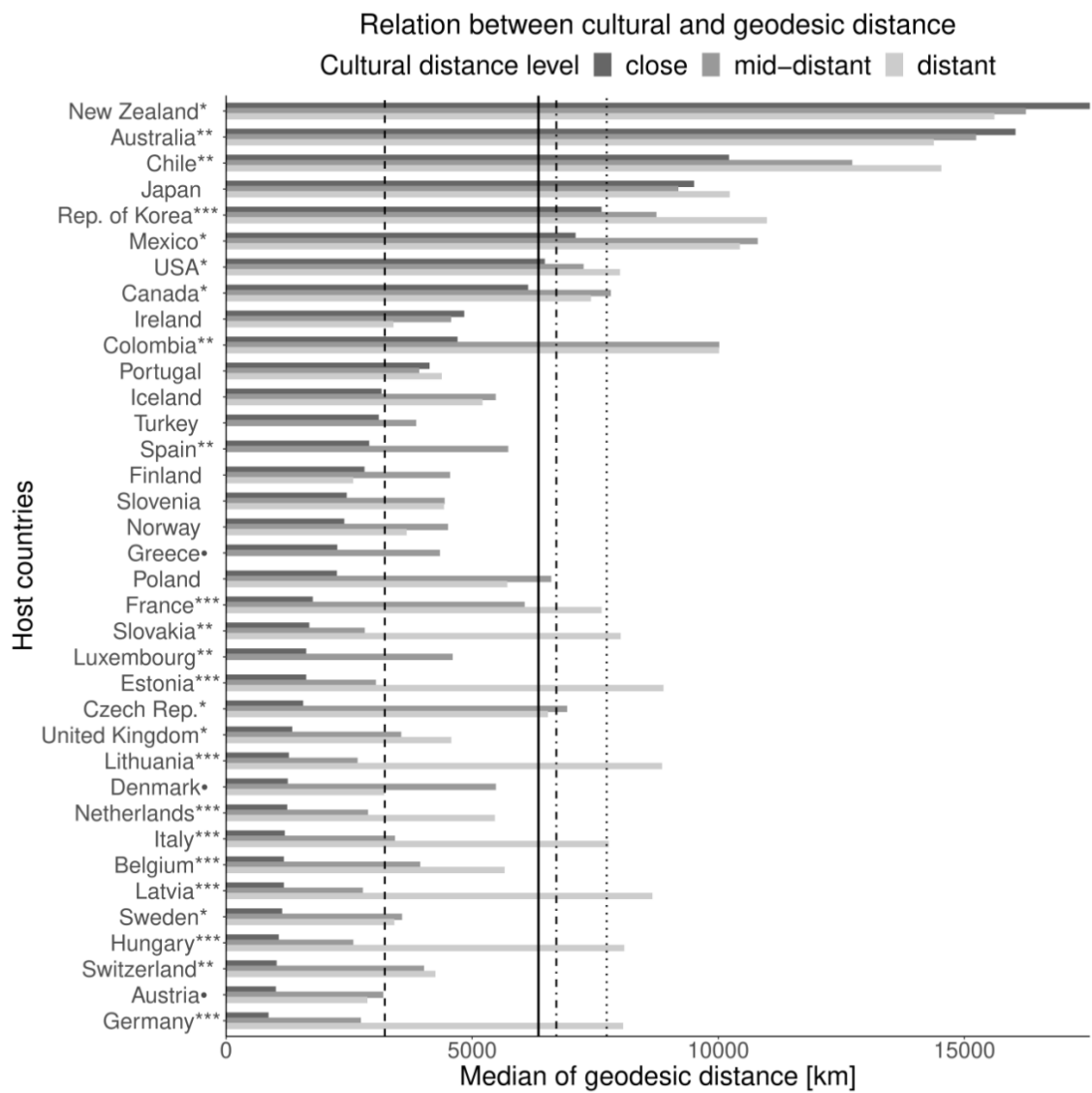

*Figure 3: Medians of geodesic distances (in kilometres) to OECD countries. Host countries are ordered according to their median geodesic distance from culturally close countries. Countries of origin are grouped by their cultural distances to each host country. The pair of countries is classified as culturally close (distant) if the cultural distance is less than 61.6 (greater than 93.4); otherwise, the pair is considered culturally mid-distant. The dashed, dash-and-dot, and dotted lines represent the median of geodesic distance from culturally close, culturally mid-distant, and culturally distant countries to OECD countries. The solid line represents the grand median of geodesic distance to OECD countries. Significant relationships between cultural and geodesic distances are indicated next to the names of the countries.*

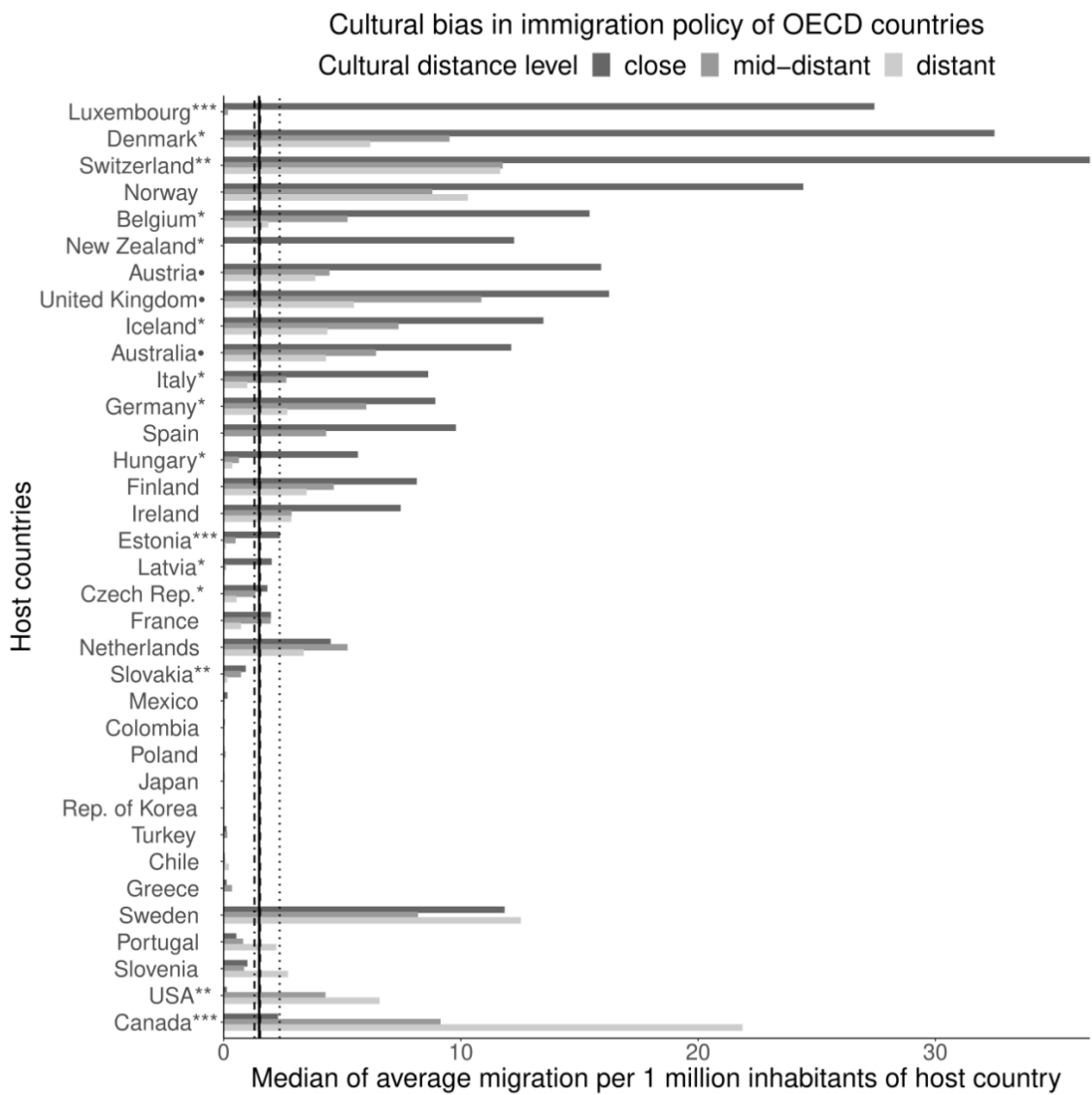

*Figure 4: Medians of average migration to OECD countries between 1995 and 2015, measured per year per 1 million inhabitants of the host country. Host countries are ordered according to differences in medians of average migration with respect to cultural distance. Countries of origin are grouped by their cultural distance to each host country. The pair of countries is classified as culturally close (distant) if the cultural distance is less than 61.6 (greater than 93.4); otherwise, the pair is considered culturally mid-distant. The dashed, dash-and-dot, and dotted lines represent the median of average migration from culturally close, culturally mid-distant, and culturally distant countries to OECD countries, respectively. The solid line represents the grand median of average migration to OECD countries. Significant bias in migration towards cultural (dis)similarity is indicated next to the names of the countries.*

Only Canada and the USA showed a significant bias towards culturally distant countries. Canada, much more than the USA, accepts immigrants from culturally different countries, increasing cultural diversity despite the possible negative outcomes stemming from it. In contrast, 16 countries showed a bias in migration patterns towards culturally close migrants. Among these countries are all Central European countries, with the exception of Poland (i.e., Switzerland, Germany, Austria, the Czech Republic, Slovakia, Hungary), one country from Southern Europe (Italy), two Scandinavian countries (Denmark and Iceland), two countries of Benelux (Belgium and Luxembourg), two of the three Baltic countries (Estonia and Latvia), Australia, New Zealand, and the United Kingdom.

Among former European colonial nations, with the notable exception of Spain, whether they were large (the United Kingdom, France, Portugal, the Netherlands) or small (Denmark, Sweden, Belgium, Norway, Germany) colonisers, we observe varying degrees of culturally distant immigrants, even if, in some cases, there are no statistically significant results. Sweden, in particular, exhibits a similar affinity to culturally close and distant migrants, making their migrant policy most similar to those of Canada and the USA. All other non-European OECD countries (Mexico, Chile, Colombia, Japan, and the Republic of Korea) have very low numbers of migrants, and are statistically insignificant. From 1995 to 2015, there was no significant migration to Lithuania from the countries considered.

Finally, let us examine the hypothesis that the average number of migrants leaving OECD countries is related to cultural similarities to the host country. In general, the grand median of the average number of migrants leaving OECD countries between 1995 and 2015 is 0.98 per year per 1 million inhabitants of the country of origin. The median of average number of migrants leaving for culturally close OECD countries is 1.37 per year per 1 million inhabitants of the origin country. For culturally mid-distant countries, the median was 0.74, and for culturally distant countries it was 1.20.

More than one-third of the countries (16 OECD countries and 18 non-OECD countries) showed significant bias, whether positive or negative, towards cultural similarity between the migrant and the potential host country (see Figure 5). Concerning OECD countries, migrants from the USA, Canada, Australia, Sweden, Iceland, the United Kingdom, Belgium, Switzerland, Germany, Austria, and Denmark prefer culturally similar host countries. In contrast, people from Spain, Portugal, Slovenia, Turkey, and the Republic of Korea migrate more to culturally different OECD countries. For the remaining 20 OECD countries, there were no significant differences in migration to other OECD countries depending on cultural distance.

Concerning the 57 non-OECD countries, Montenegro and Puerto Rico did not have significant migration to OECD countries during the study period. Only two countries, South Africa and Trinidad and Tobago, exhibit a significant preference for migration to culturally similar OECD countries. In contrast, a tendency towards migration to culturally distant OECD countries was observed in more than one-quarter of the non-OECD countries (16 countries: Bangladesh, Bosnia and Herzegovina, Croatia, Egypt, Ghana, Indonesia, Iran, Iraq, Jordan, Lebanon, Nigeria, Peru, Saudi Arabia, Thailand, Vietnam, and Zambia). For almost two-thirds of non-OECD countries (37 out of 57), there are no significant differences in migration to OECD countries depending on cultural distance.

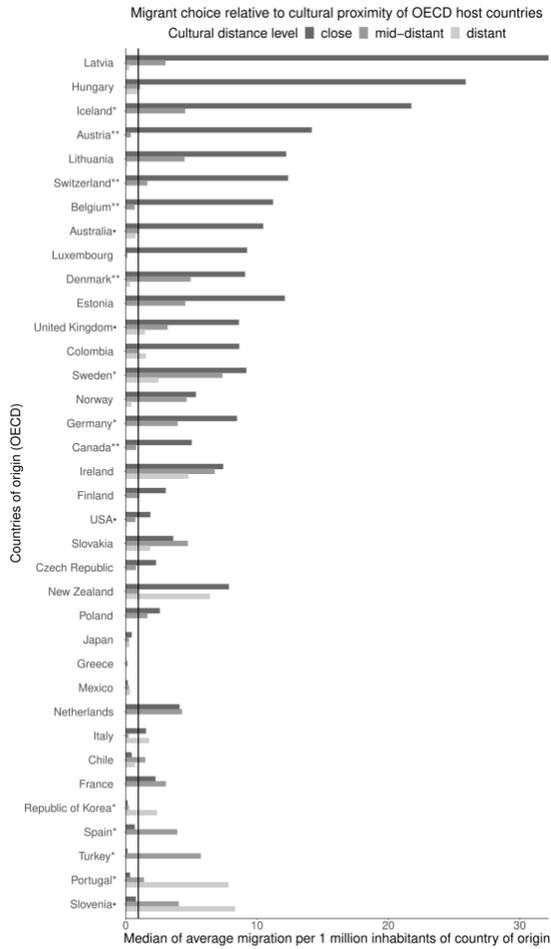

*Figure 5a: OECD countries of origin*

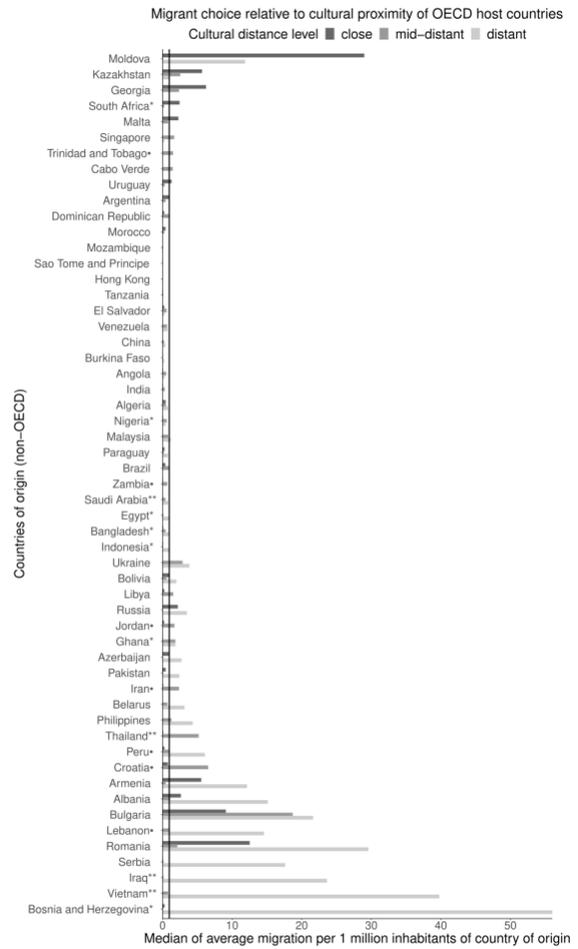

*Figure 5b: non-OECD countries of origin*

*Figure 3: Medians of average immigration to OECD countries between 1995 and 2015, measured per year per 1 million inhabitants of the country of origin. Countries of origin are ordered according to differences in medians of average immigration with respect to cultural distance. The host countries are grouped by their cultural distance to each country of origin. The pair of countries is classified as culturally close (distant) if the cultural distance is less than 61.6 (greater than 93.4); otherwise, the pair is considered culturally mid-distant. The solid line represents the grand median of average immigration to OECD countries. Significant bias in immigration towards cultural (dis)similarity is indicated next to the names of the countries.*

# 5 Conclusions

Our results clearly suggest that culture has a more significant influence on migration patterns in some countries than in others.

First, national culture makes some countries close, whereas others distant. As expected, the median of geodesic distance increases with increasing cultural distance and vice versa in the majority of countries. However, for a small number of countries, New Zealand and Australia, in particular, cultural differences decrease with increasing geodesic and vice versa. Moreover, we found some OECD countries, namely Colombia, Denmark, and Japan, which have cultures that are distant, even from countries with minimal geodesic distance.

Second, splitting the OECD countries into close and distant relative to the migrants' country of origin also allowed us to study the migration pattern from the migrants' perspective. Our results show that some migrants from relatively rich countries choose host countries based on their proximity to their own culture. On the other hand, migrants from poorer countries migrate to culturally distant countries several times more often than to culturally close ones.

Finally, as high cultural diversity is associated with many disadvantages for the host country, about half of the OECD countries (16 out of 38) showed a statistically significant bias towards the acceptance of migrants culturally close to them. The most biased in this regard are Switzerland, Luxembourg, Denmark, Belgium, New Zealand, Hungary, Estonia, and Latvia. A mild bias towards culturally close immigrants can be found in Austria, the United Kingdom, Iceland, Australia, Italy, Germany, the Czech Republic, and Slovakia. Migrants from culturally distant countries predominated in only two OECD countries: Canada and the USA. The remaining countries showed no bias or no statistically significant results. Here, the question arises: to what extent are these results related to the immigration policies of OECD countries?

This study attempted to capture migration patterns from a long-term perspective. Owing to the availability of data and policy changes in Central and Eastern Europe, the OECD data and migration period between 1995 and 2015 were chosen for the analysis. However, the study is partially limited by the choice of countries of origin included in the study. At present, data on all six cultural dimensions are available only for the 93 countries mentioned.

# Declarations

**Ethics approval and consent to participate:** Not applicable.

**Consent for publication:** Not applicable.

**Availability of data and material:** The data that support the findings of this study are openly available in the United Nations database, in the official Hofstede Insight's Cultural Executive Ownership Program website at www.hofstede-insights.com/country-comparison, and in the World Bank database at data.worldbank.org/indicator/SP.POP.TOTL. The datasets used and analysed during the current study are available from the corresponding author upon reasonable request.

**Competing interests:** The authors declare that they have no competing interests.

**Funding:** This research was supported by the grant IGA_PrF_2021_008 Mathematical Models of the Internal Grant Agency of Palacký University Olomouc.

**Authors' contributions:** TE conducted the literature review, conceptualised the article, and wrote the manuscript; EF conducted the statistical analysis and wrote the manuscript; AE prepared the data and conducted the statistical analysis. All authors have read and approved the final manuscript.

**Acknowledgements:** The authors thank Tobiáš Unger for helpful support with the technical preparation of the manuscript and for valuable discussions on the topic. His contributions have improved the quality of this study.